\documentclass[12pt]{article}
\usepackage{a4wide}
\usepackage{cite}
\usepackage{amsmath}
\usepackage{amstext,amsfonts,amsbsy,eucal,amssymb}
\usepackage{amsmath,pifont}
\usepackage{color}
\usepackage{xypic}
\xyoption{all}
\xyoption{knot}
\usepackage{epsfig,psfrag}

\numberwithin{equation}{section}
\renewcommand{\thefootnote}{\fnsymbol{footnote}}

\def\openone{\leavevmode\hbox{\small1\kern-3.8pt\normalsize1}}%
\DeclareMathOperator{\sh}{sh}
\DeclareMathOperator{\ch}{ch}

\renewcommand{\Im}{\text{Im}}

\newcommand{\Li}{\mathrm{Li}_2}
\newcommand{\VCS}{\mathrm{VCS}}
\newcommand{\CS}{\mathrm{CS}}
\newcommand{\Vol}{\mathrm{Vol}}

\begin{document}

\baselineskip 21pt
\parskip 7pt

\noindent

\hfill Revised on July 14, 2001

\vspace{24pt}

\begin{center}

  {\Large\textbf{
      Hyperbolicity of Partition Function
      and Quantum Gravity
      }
    }

  \vspace{24pt}

  {\large Kazuhiro \textsc{Hikami}}
  \footnote[2]{
    \texttt{hikami@phys.s.u-tokyo.ac.jp}
    }

  \vspace{2pt}
   \textsl{Department of Physics, Graduate School of Science,\\
     University of Tokyo,\\
     Hongo 7--3--1, Bunkyo, Tokyo 113--0033, Japan.
     }

(Received: \hspace{40mm})

\vspace{18pt}

\noindent
\underline{\textsf{ABSTRACT}}
\end{center}

We study a geometry of the partition function which is
defined in terms of a solution of the
five-term relation.
It is shown  that  the 3-dimensional
hyperbolic structure or Euclidean AdS$_3$
naturally arises in the
classical limit of this invariant.
We discuss
that the oriented  ideal tetrahedron can
be assigned to the partition function of string.

\vfill
\noindent


\newpage

\renewcommand{\thefootnote}{\arabic{footnote}}
\section{Introduction}

This article is devoted to reveal a relationship between the partition
function and 3-dimensional hyperbolic geometry.
From the physical viewpoint,
the 3-dimensional hyperbolic space $\mathbb{H}^3$ corresponds to the
Euclidean Anti-de Sitter space AdS$_3$, and
many studies concerning
the  AdS/CFT correspondence~\cite{Malda98a} including
string theory on AdS and
the $SL(2,\mathbb{R})$ WZNW model
have been done
(for a review, see Ref.~\citen{AhaGubOogYOz00a} and references therein).
Our motivation is based on  a recent conjecture 
that an asymptotic behavior of a specific value of
the colored Jones polynomial gives the hyperbolic volume of knot
complement~\cite{Kasha95,Kasha96b,MuraMura99a,YYokot00b,Hikam00d,BaseiBened01a}.
This opens up a geometrical study of the quantum
knot invariants such as the Jones polynomial, which have been
introduced based on the quantum group.
Here 
we shall give a geometrical picture of the partition function which
indicates an aspect of the AdS/CFT correspondence;
we show  that the bosonic and fermionic partition function
constitutes as oriented ideal tetrahedron in $\mathbb{H}^3$.
This partition function can be seen as a string partition function on
knot, which is a collection of torus and is on the boundary of $\mathbb{H}^3$.

The organization of this paper is as follows.
In section~\ref{sec:space} we briefly review a fact of the hyperbolic
space $\mathbb{H}^3$.
Almost all orientable  3-manifolds are believed to admit a hyperbolic
structure, and the hyperbolic volume of manifold as invariant
can be constructed based on the ideal triangulation.
In section~\ref{sec:five}
we study  several properties of the  quantum dilogarithm function.
Key identity is the five-term relation, and we
define the partition function by assigning oriented tetrahedron to the
quantum dilogarithm function.
In section~\ref{sec:classical}
we study a classical limit of the partition function.
We show that the oriented tetrahedron can be regarded as the ideal
tetrahedron in $\mathbb{H}^3$, and that
the hyperbolicity consistency  condition in gluing tetrahedra
exactly coincides with the critical point of the
partition function.
In section~\ref{sec:physics}
we consider an application in physics, and
we clarify a relationship with the AdS/CFT correspondence.
As was pointed out in Refs.~\citen{LFadd95a,LFadd99b}, the quantum
dilogarithm
function used here is a modular double of the  $q$-exponential
function, which coincides with the partition function of string from cylinder.
We show that
the chemical potential and the
average number of angular momentum
determines the ideal tetrahedron.
The last section is for concluding remarks.

\section{Hyperbolic Space $\mathbb{H}^3$}
\label{sec:space}

The 3-dimensional hyperbolic space $\mathbb{H}^3$
(see \emph{e.g.} Refs.~\citen{WPThurs80Lecture,Ratcl94a}),   
or the Euclidean version of Anti-de Sitter space AdS$_3$,
is defined as the
space-like surface in
4-dimensional Minkowski space;
\begin{align*}
  &-x_0^{~2} + x_1^{~2} + x_2^{~2} + x_3^{~2} = -1,
  &
  x_0 > 0 .
\end{align*}
Here the metric is
$\mathrm{d} s^2
=
- \mathrm{d} x_0^{~2} +\mathrm{d}x_1^{~2}
+\mathrm{d}x_2^{~2} +\mathrm{d}x_3^{~2}
$,
and it has a constant negative curvature $-1$.
The hyperbolic space $\mathbb{H}^3$ is conformally mapped into the
3-dimensionally disk $\mathbb{D}_3$
(the Poincar{\'e} model)
via
\begin{gather*}
   x \mapsto \frac{(x_1, x_2, x_3)}{1+x_0},
  \\[2mm]
  \left\{
  \begin{array}{ll}
    x_0  = \ch \xi,
    &
    x_1 = \sh \xi \sin \theta \, \cos \varphi ,
    \\[2mm]
    x_2  = \sh \xi \sin \theta \sin \varphi,
    &
    x_3 = \sh \xi \cos \theta
  \end{array}
  \right.
\end{gather*}
with
$\mathrm{d}s^2=\mathrm{d} \xi^2 + \sh^2 \xi \,
( \mathrm{d} \theta^2 + \sin^2 \theta \, \mathrm{d} \varphi^2)$,
and boundary $\partial \mathbb{H}^3$
of the hyperbolic space is homeomorphic to $S^2$.
The orientation-preserving isometry of $\mathbb{H}^3$ is isomorphic to
$PSL(2,\mathbb{C})
$.
Hereafter to consider the ideal tetrahedron, we use the Beltrami
half-space model $X_0>0$ with coordinate $(X_0, X_1, X_2)$,
and the metric is given by
\begin{equation*}
  \mathrm{d}s^2
  =
  \frac{\mathrm{d} X_0^{~2} + \mathrm{d} X_1^{~2} + \mathrm{d} X_2^{~2}}{X_0^{~2}} .
\end{equation*}
The transformation is explicitly given by
\begin{align*}
  X_0^{~-1}
  & = \ch \xi + \sh \xi \cos \theta,
  &
  X_1
  &= X_0 \sh \xi \sin \theta \cos \varphi,
  &
  X_2
  &= X_0 \sh \xi \sin \theta \sin \varphi .
\end{align*}
With this metric
geodesics are semi-circles orthogonal to
$\{ X_0=0  \} \cup \infty$.

Every non-compact finite volume hyperbolic 3-manifold admits a
decomposition into a finite number of ideal
polyhedra~\cite{WPThurs80Lecture,EpstPenn88a}.
The  ideal polyhedron means that
all vertices lie on
$\partial \mathbb{H}^3$, and all of edges are hyperbolic geodesics.
Among them ideal tetrahedron
is completely determined by a single complex
number $z$ with positive imaginary part,
which we call modulus, or cross-ratio;
this follows from that the ideal tetrahedron is determined by the
three dihedral angels $\alpha$, $\beta$, and $\gamma$
of edges satisfying $\alpha+ \beta+\gamma=\pi$.
For fixed edge $e$,
the  parameter
which describes the dihedral angle
(see Fig.~\ref{fig:triangle})
is given in terms of modulus by 
\begin{equation*}
  z_i(e) \in
  \left\{
    z_i , \frac{1}{1-z_i} , 1 - \frac{1}{z_i}
  \right\} .
\end{equation*}
The hyperbolic volume of the ideal tetrahedron is
given by  the Bloch--Wigner function $D(z)$
(see eq.~\eqref{define_Bloch_Wigner})~\cite{JMiln82a}.
Depending on the orientation of tetrahedron,
we can extend the modulus 
to  $z \in \mathbb{C}$.

\begin{figure}[htbp]
  \centering
  \begin{equation*}
    \xy/r1.0pc/:.
    {
      { (0,0) \ar@{-} (10,0) },
      { (0,0) \ar@{-} (6,8) },
      { (10,0) \ar@{-} (6,8) },
      { (-0.5,-0.5)*+{0} },
      { (10.5,-0.5)*+{1} },
      { (6,8.5)*+{z} },
      { (1,0.5)*+{z} },
      { (8.2,1)*+{\frac{1}{1-z}} },
      { (5.7,5.8)*+{{1-\frac{1}{z}}} },
      }
    \endxy
  \end{equation*}
  \caption{}
  \label{fig:triangle}
\end{figure}

Gluing a finite collection of
ideal tetrahedra
together results in a 3-manifold $M$ admitting a hyperbolic structure of
finite volume~\cite{WPThurs80Lecture}.
To endow the hyperbolic structure in $M$,
the following gluing conditions must be satisfied;
\begin{enumerate}
\item 
  triangles cut out of adjacent
  tetrahedra to every edge $e_\nu$
  to 
  fill  neatly around edge,
  \emph{i.e.},
  we  require the consistency condition
  \begin{align*}
    & \sum_j \arg z_j(e_\nu) = 2 \, \pi,
    &
    & \prod_j z_j(e_\nu) = 1 .
  \end{align*}

\item
  the developing map near the ideal vertex
  yields a Euclidean structure on the horosphere.

%
\end{enumerate}
\emph{i.e.}
an ideal triangulation of a 3-manifold $M$, when $\partial M$ is a
collection of tori,
$M-\partial M$  have a hyperbolic structure if and only if for each
edge the hyperbolic consistency condition is fulfilled.
It is not   clear for a  case of closed manifold,
which
can be obtained topologically by Dehn
filling a certain hyperbolic link complement.
When a 3-manifold $M$ is triangulated in this way,
a sum of the hyperbolic volume of the ideal tetrahedron
\begin{equation}
  \label{volume_invariant}
  \Vol(M) = \sum_i D(z_i) 
\end{equation}
is a topological invariant of $M$.

The
volume of oriented hyperbolic 3-manifold $M$ has
an analytic relationship with the Chern--Simons invariant
$\CS(M)$~\cite{NeumZagi85a,TYoshi85a}.
Namely
\begin{equation}
  \label{define_v_cs}
  \VCS(M)
  = \Vol(M) + \mathrm{i} \, \CS(M)
\end{equation}
is a natural complexification of $\Vol(M)$,
and in general the formulae to compute
$\CS(M)$ give $\Vol(M)$.
The invariant  $\VCS(M)$ is induced from the pre-Bloch
group~\cite{WDNeum92a} by
\begin{equation}
  \label{represent_VCS}
  \begin{array}{lcccc}
    \mathrm{i} \,\VCS 
    & :   & \mathcal{P}(\mathbb{C})  & \longrightarrow & \mathbb{C}
    \\[2mm]
    & & [z_i] & \longmapsto & L(1-z_i) 
  \end{array}
\end{equation}
Here $[z]$ is the element of  the Bloch group satisfying
\begin{equation}
  \label{Bloch_group}
  [x] - [y] + [\frac{y}{x}]
  -[\frac{1-x^{-1}}{1-y^{-1}}]
  +[\frac{1-x}{1-y}] = 0 ,
\end{equation}
and $L(z)$ is the Rogers dilogarithm function~\eqref{define_Rogers}.
Fact that eqs.~\eqref{volume_invariant} and ~\eqref{define_v_cs} are
invariant of $M$ is based on that
the Bloch invariant defined by
\begin{equation*}
  \beta(M) = \sum_i [ z_i ]
\end{equation*}
is the invariant of the manifold~\cite{NeumYang99a}.


\section{Quantum Dilogarithm Function and Partition Function}
\label{sec:five}

We study the  integral,
\begin{equation}
  \label{define_Phi}
  \Phi_\gamma(\varphi)
  =
  \exp
  \left(
    \int_{\mathbb{R} + \mathrm{i} 0}
    \frac{
      \mathrm{e}^{- \mathrm{i} \, \varphi \, x}
      }{
      4 \sh(\gamma \, x) \, \sh(\pi \, x)
      }
    \frac{\mathrm{d}x}{x}
  \right) ,
\end{equation}
which was first introduced by Faddeev~\cite{LFadd95a}.
As we will discuss later, this function was originally introduced as a
dualization or modular double of the $q$-exponential function.
The function enjoys the five-term
relation~\cite{ChekFock99a,FaddKashVolk00a},
\begin{equation}
  \label{pentagon_Phi}
  \Phi_\gamma(\Hat{p}) \, \Phi_\gamma(\Hat{q})
  =
  \Phi_\gamma(\Hat{q}) \, \Phi_\gamma(\Hat{p} + \Hat{q}) \,
  \Phi_\gamma(\Hat{p})  ,
\end{equation}
where $\Hat{p}$ and $\Hat{q}$ are the canonically conjugate operators
satisfying
\begin{equation}
  \label{canonical}
  [ \Hat{p} ~,~ \Hat{q} ] = - 2 \, \mathrm{i} \, \gamma  .
\end{equation}
The five-term relation~\eqref{pentagon_Phi} is rewritten into simple
form,
\begin{equation}
\label{five_term_S}
  S_{2,3} \, S_{1,2}
  = S_{1,2} \, S_{1,3} \, S_{2,3} ,
\end{equation}
where the $S$-operator is defined by
\begin{equation}
  S_{1,2}
  =\mathrm{e}^{\frac{1}{2\mathrm{i}\gamma} \, \Hat{q}_1 \, \Hat{p}_2}
  \cdot
  \Phi_\gamma ( \Hat{p}_1 + \Hat{q}_2 - \Hat{p}_2) .
\end{equation}
Here $\Hat{p}_1=\Hat{p}\otimes \openone$,
$\Hat{q}_2 = \openone \otimes \Hat{q}$, and so on.
The  matrix element of the $S$-operator over the momentum
space,
$
  \Hat{p} \ | \ p \rangle
  =
  p \ | \ p \rangle 
$, is given by~\cite{Hikam00d}
\begin{equation}
  \label{represent_S}
  \begin{aligned}
    \langle p_1, p_2 \ | \ S \ | \ p_1^\prime , p_2^\prime \rangle
    =
    \delta(p_1 + p_2 - p_1^\prime) \cdot
    \Phi_\gamma(p_2^\prime - p_2 + \mathrm{i} \pi + \mathrm{i} \gamma)
    \cdot
    \mathrm{e}^{\frac{1}{2\mathrm{i} \gamma}
      \left(
        -\frac{\pi^2 + \gamma^2}{6} - \frac{\gamma \pi}{2}
        + p_1 (p_2^\prime - p_2)
      \right)
      } ,
    \\[2mm]
    \langle p_1, p_2 \  | \ S^{-1} \ | \ p_1^\prime , p_2^\prime \rangle
    =
    \delta(p_1 - p_1^\prime - p_2^\prime) \cdot
    \frac{1}{
      \Phi_\gamma(p_2 - p_2^\prime - \mathrm{i} \pi - \mathrm{i} \gamma)
      }
    \cdot
    \mathrm{e}^{\frac{1}{2\mathrm{i} \gamma}
      \left(
        \frac{\pi^2 + \gamma^2}{6} + \frac{\gamma \pi}{2}
        - p_1^\prime (p_2 - p_2^\prime)
      \right)
      } .
  \end{aligned}
\end{equation}

\begin{figure}[thbp]
  \begin{center}
        \parbox{2cm}
    {$\langle p_1, p_2 \ | \ S  \ | \ p_1^\prime , p_2^\prime \rangle ~:~$}
    \begin{psfrags}
      \psfrag{A}{\textcolor{red}{\reflectbox{$p_1$}}}
      \psfrag{B}{\rotatebox{40}{$p_2$}}
      \psfrag{C}{\rotatebox{-10}{$p_1^\prime$}}
      \psfrag{D}{\textcolor{red}{\reflectbox{$p_2^\prime$}}}
      \psfrag{a}{$z_1$}
      \psfrag{b}{$z_2$}
      \psfrag{c}{$z_3$}
      \psfig{file=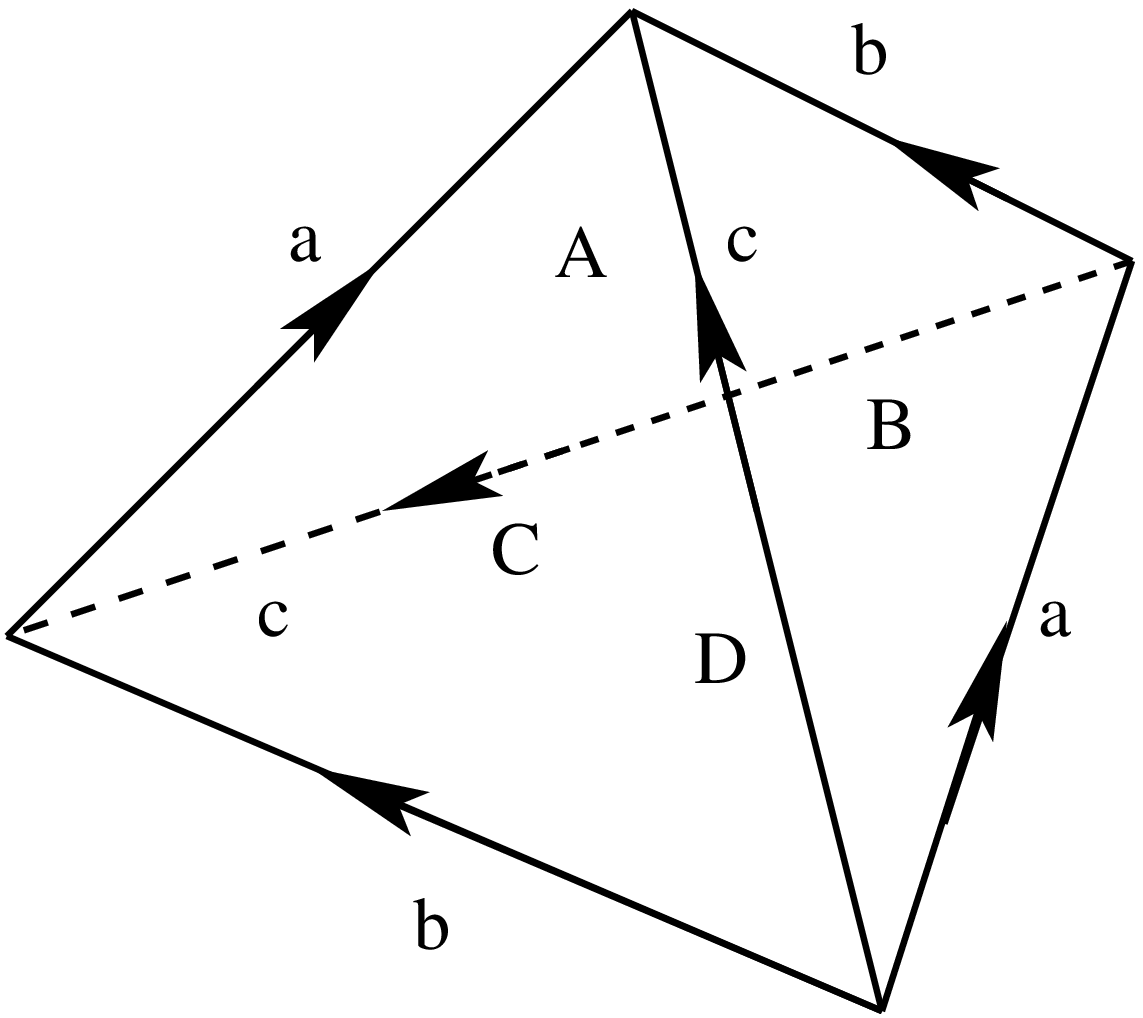,scale=0.4}
    \end{psfrags}
    \hspace{20mm}
    \parbox{2cm}
    {$\langle p_1, p_2 \ | \ S^{-1} \ | \ p_1^\prime , p_2^\prime \rangle ~:~$}
    \begin{psfrags}
      \psfrag{A}{\textcolor{red}{\reflectbox{$p_1$}}}
      \psfrag{B}{\textcolor{red}{\reflectbox{$p_2$}}}
      \psfrag{C}{\rotatebox{-10}{$p_1^\prime$}}
      \psfrag{D}{\rotatebox{40}{$p_2^\prime$}}
      \psfrag{a}{$z_1$}
      \psfrag{b}{$z_2$}
      \psfrag{c}{$z_3$}
      \psfig{file=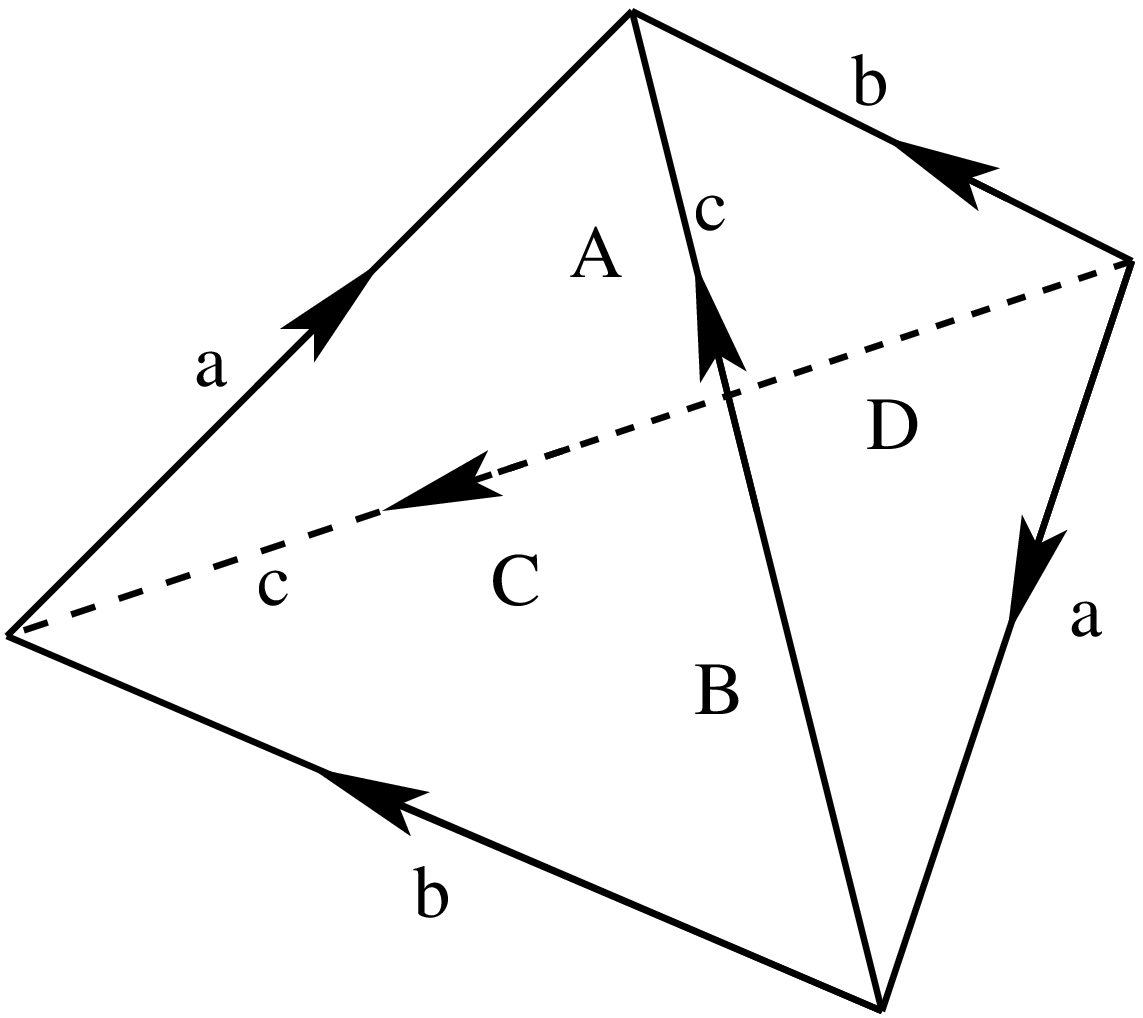,scale=0.4}
    \end{psfrags}
    \caption{
      The $S$-operators~\eqref{represent_S} denote
      the oriented tetrahedron.
      Parameters $p$ are assigned to every face, and $z_a$ denotes a
      dihedral angle.
      In the classical limit $\gamma \to 0$, those  become
      hyperbolic       ideal tetrahedra with
      modulus $\mathrm{e}^{p_2^\prime - p_2}$,
      and the dihedral angles $z_a$ at the edges of the simplex
      are 
      $z_1=\mathrm{e}^{p_2^\prime -p_2}$,
      $z_2=1-z_1^{~-1}$, and
      $z_3=(1-z_1)^{-1}$, respectively,
      with opposite edges having the same angle.
      }
    \label{fig:tetrahedron}
  \end{center}
\end{figure}

\begin{figure}[htbp]
  \begin{center}
    \psfig{file=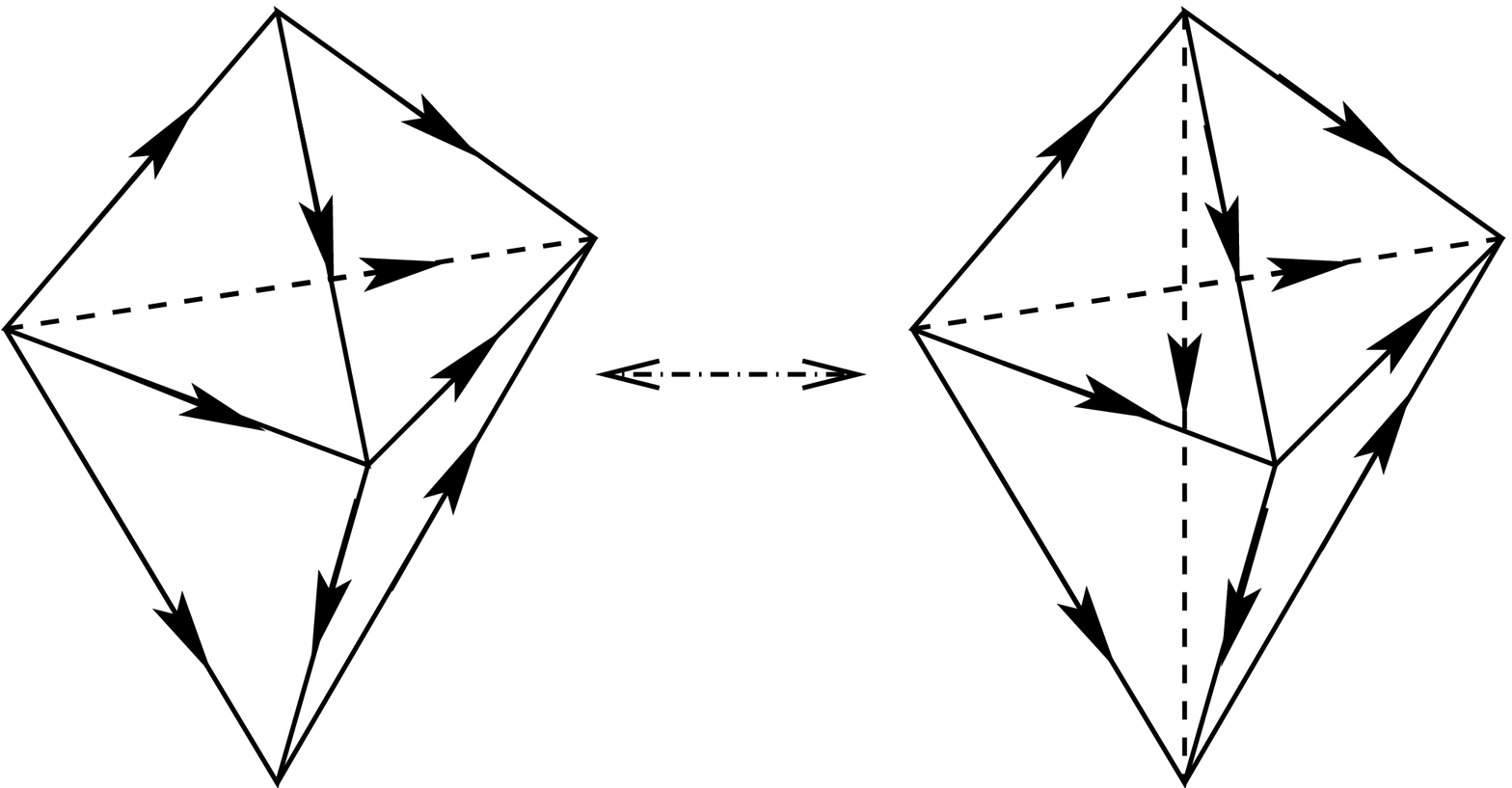,scale=0.5}
    \caption{
      Pentagon identity~\eqref{five_term_S}.
      }
    \label{fig:pentagon}
  \end{center}
\end{figure}

We can give the geometrical picture for this operator;
a key is the five-term relation~\eqref{five_term_S}.
We associate the (3-dimensional) oriented tetrahedron to the
$S$-operators;
each 2-simplex is assigned the momentum
(Fig.~\ref{fig:tetrahedron}).
With this identification, we can glue two 2-simplex in pairs so that
the orientation of edges match.
For instance, the
pentagon relation~\eqref{five_term_S}
can be depicted as Fig.~\ref{fig:pentagon}.

When 3-manifold $M$ is decomposed into  ideal tetrahedron,
we define the partition function $\tau(M)$ 
by
\begin{equation}
  \label{partition_function}
  \tau(M)
  =
  \iint \mathrm{d} {p} \
  \prod
  \langle p_i , p_j  \ | \ S^{\pm 1} \ | \ {p}_k , p_l \rangle .
\end{equation}
In the following we study in detail the classical limit of the
partition function $\tau(M)$.

\section{Classical Limit and Hyperbolic Geometry}
\label{sec:classical}

We study  the classical limit $\gamma\to 0$ of matrix
elements~\eqref{represent_S}
of the
$S$-operators.
As simple calculation leads that the Faddeev integral reduces to
the Euler dilogarithm function~\eqref{define_Euler}
\begin{equation}
  \label{Phi_and_Euler}
  \Phi_\gamma(x) = \exp \frac{1}{2 \, \mathrm{i} \, \gamma} \,
  \Bigl(
  \Li(- \mathrm{e}^{x})
  + O(\gamma)
  \Bigr) ,
\end{equation}
we have
\begin{equation}
  \label{asymptotic_S}
  \begin{aligned}
  \langle p_1 , p_2 \ | \ S \ | \ p_1^\prime , p_2^\prime \rangle
  =
  \delta(p_1+p_2 - p_1^\prime) \cdot
  \exp
  \left(
    -\frac{1}{2 \,  \mathrm{i} \, \gamma}
    V(p_2^\prime - p_2 , p_1)
    + O(\gamma^0)
  \right) ,
  \\[2mm]
  \langle p_1 , p_2 \ | \ S^{-1} \  | \ p_1^\prime , p_2^\prime \rangle
  =
  \delta(p_1-p_1^\prime - p_2^\prime) \cdot
  \exp
  \left(
    \frac{1}{2 \,  \mathrm{i} \, \gamma}
    V(p_2 - p_2^\prime , p_1^\prime)
    + O(\gamma^0)
  \right) ,
\end{aligned}
\end{equation}
where
\begin{equation}
  V(x,y)
  =
  \frac{\pi^2}{6}
  - \Li(\mathrm{e}^x) - x \, y .
\end{equation}
Thus 
the operators  $S^{\pm 1}$, which  is assigned to oriented 
tetrahedron in the
partition function~\eqref{partition_function},
asymptotically reduces to the  function
$V(x,y)$ satisfying
\begin{gather}
  \label{function_V_L}
  V(x,y)
  =
  L(1 - \mathrm{e}^x)
  +\frac{1}{2}
  \left(
    \frac{\partial V(x,y)}{\partial \log x}
    +
    \frac{\partial V(x,y)}{\partial \log y}
  \right) ,
  \\[2mm]
  \Im \, V(x,y)
  =
  D(1 - \mathrm{e}^x)
  + \log | \mathrm{e}^x | \cdot
  \Im \, \left(
    \frac{\partial V(x,y)}{\partial x}
  \right)
  + \log | \mathrm{e}^y | \cdot
  \Im \, \left(
    \frac{\partial V(x,y)}{\partial y}
  \right) .
\end{gather}
Here we have used  the Rogers dilogarithm~\eqref{define_Rogers} and
the Bloch--Wigner function~\eqref{define_Bloch_Wigner}.
As  the Bloch--Wigner function $D(z)$ gives the
hyperbolic volume of the ideal tetrahedron~\cite{JMiln82a}, we are
forced to
study  a relationship between
the $S$-operator at the critical point
and the hyperbolic geometry.

For our purpose
we reconsider the five-term
relation~\eqref{five_term_S} in the classical limit in detail.
We substitute the asymptotic form~\eqref{asymptotic_S} into
matrix form of eq.~\eqref{five_term_S},
$\langle p_1 , p_2 , p_3 \ | \ S_{2,3} \, S_{1,2} \  | \ p_1^\prime ,
p_2^\prime, p_3^\prime \rangle
=
\langle p_1 , p_2 , p_3 \ | \ S_{1,2} \,S_{1,3} \, S_{2,3}
| \ p_1^\prime , p_2^\prime,
p_3^\prime \rangle
$.
LHS reduces to
\begin{multline}
  \label{pentagon_LHS}
  \int \mathrm{d} x \
  \langle p_2 , p_3 \ | \ S \  | \  x , p_3^\prime \rangle \,
  \langle p_1 , x \ | \ S \  | \ p_1^\prime , p_2^\prime \rangle
  \\
  =
  \delta(p_1 + p_2 + p_3 - p_1^\prime) \cdot
  \exp
  \frac{1}{2 \, \mathrm{i} \,  \gamma} \,
  \Bigl(
    - V(p_3^\prime - p_3 , p_2)
    - V(p_2^\prime - p_2 - p_3 , p_1)
    +O(\gamma)
  \Bigr) ,    
\end{multline}
while RHS gives
\begin{multline}
  \label{pentagon_RHS}
  \iiint \mathrm{d} y \  \mathrm{d} z \  \mathrm{d} w \
  \langle p_1 , p_2 \ | \ S \  | \  y, z \rangle \,
  \langle y , p_3 \ | \ S  \ | \ p_1^\prime, w \rangle \,
  \langle z, w \ | \ S \  | \  p_2^\prime , p_3^\prime \rangle
  \\
  =
  \delta (p_1 + p_2 + p_3 - p_1^\prime) \cdot
  \int \mathrm{d} z \
  \exp \frac{1}{2 \, \mathrm{i} \, \gamma}
  \biggl(
  - \frac{\pi^2}{2}
  + \Li(\mathrm{e}^{z-p_2})
  + \Li(\mathrm{e}^{p_2^\prime - p_3 - z})
  + \Li(\mathrm{e}^{p_3^\prime - p_2^\prime + z})
  \\
  + z \,  \bigl( -p_2 + p_3^\prime - p_2^\prime + z \bigr)
  - p_1 \, p_2
  + (p_2^\prime - p_3) \, (p_1 + p_2)
  +O(\gamma)
  \biggr) .
\end{multline}
Above integral w.r.t. $z$ is evaluated by use of the saddle point method,
whose critical point is given by
\begin{align}
  (1-\mathrm{e}^{w - p_3})^{-1} \cdot
  ( 1 - \mathrm{e}^{p_2 - z}) \cdot
  (1 - \mathrm{e}^{w - p_3^\prime})
  & = 1 ,
  &
  w  & = p_2^\prime - z .
  \label{saddle_pentagon}
\end{align}
In fact,  substituting a solution
\begin{equation*}
  \mathrm{e}^{-z}
  =
  \mathrm{e}^{-p_2} + \mathrm{e}^{-p_2^\prime+p_3^\prime}
  -
  \mathrm{e}^{-p_2 - p_3 + p_3^\prime} ,
\end{equation*}
into the integral~\eqref{pentagon_RHS}, and
equating both hands sides, eq.~\eqref{pentagon_LHS}
and eq.~\eqref{pentagon_RHS} at critical point,
we get
\begin{multline*}
  \Li(\mathrm{e}^{z-p_2}) + \Li(\mathrm{e}^{p_2^\prime - p_3 - z})
  +
  \Li(\mathrm{e}^{p_3^\prime - p_2^\prime + z})
  + (z-p_2) \, ( z - p_2^\prime + p_3^\prime)
  -\frac{\pi^2}{6}
  \\
  =
  \Li(\mathrm{e}^{p_3^\prime - p_3})
  +
  \Li(\mathrm{e}^{p_2^\prime - p_2 - p_3}) .
\end{multline*}
which is nothing but
Schaeffer's five-term relation of the Euler dilogarithm
function,
\begin{equation*}
  \Li(\frac{1-x^{-1}}{1-y^{-1}})
  =
  \Li(x) - \Li(y) + \Li(\frac{y}{x})
  +
  \Li(\frac{1-x}{1-y})
  -\frac{\pi^2}{6}
  + \log x \log (\frac{1-x}{1-y}) ,
\end{equation*}
with $x=\mathrm{e}^{z-p_2}$
and
$y=\mathrm{e}^{p_2^\prime - p_2 - p_3}$.

A meaning of the saddle point equation~\eqref{saddle_pentagon}
becomes clear once
we regard  the oriented
tetrahedron in Fig.~\ref{fig:tetrahedron}
as  a hyperbolic
ideal one;
the modulus is
$\mathrm{e}^{p_2^\prime - p_2}$,
and
the dihedral angles $z_a$ at the edges of the simplex
are 
\begin{align}
  z_1 &=\mathrm{e}^{p_2^\prime -p_2},
  &
  z_2 & =1-z_1^{~-1} ,
  &
  z_3 & =(1-z_1)^{-1},
\end{align}
with opposite edges having the same angle
(see e.g. Ref.~\citen{WPThurs80Lecture}).
Then the condition~\eqref{saddle_pentagon}
exactly coincides with the hyperbolic consistency condition for gluing
three tetrahedra around central edge of RHS in
Fig.~\ref{fig:pentagon}.

The correspondence between the saddle point condition and the
hyperbolic consistency condition can be checked
for
other forms of the five-term relations
such as
$S_{2,3} \, S_{1,2} \, S_{2,3}^{-1} = S_{1,2} \, S_{1,3}$,
and  a trivial identity
$S_{1,2} \, S_{1,2}^{~-1} = 1$.
See Ref.~\citen{Hikam00d} for further correspondence between the
saddle point equation and the consistency condition, which appeared in
an ideal triangulation of the knot complement.
As a result
the oriented tetrahedron, which we have assigned to
the $S$-operator based on the
five-term relation~\eqref{pentagon_Phi},
is identified with the ideal tetrahedron in the classical limit.
Correspondingly
the partition function of 3-manifold $M$ defined in
eq.~\eqref{partition_function}
reduces to
\begin{equation}
  \label{volume_conjecture}
  \lim_{\gamma \to 0}
  2 \, \mathrm{i} \, \gamma
  \log \tau(M)
  =
  \sum_i L(1-\mathrm{e}^{z_i}) ,
\end{equation}
which
may give the hyperbolic volume and the Chern--Simons term
of $M$.

We note that the function $V(x,y)$ satisfies a one parameter
deformation of eq.~\eqref{Bloch_group};
\begin{equation*}
 V(\log x , p_1) + V( \log (\frac{y}{x}), p_1 + p_2)
 + V(\log (\frac{1-x}{1-y} ), z)
 =
 V(\log ( \frac{1-x^{-1}}{1-y^{-1}}) , p_2 )
 + V(\log y , p_1) ,
\end{equation*}
where $x$, $y$, and $z$ are given above in terms of $p_2$, $p_3$,
$p_2^\prime$, and $p_3^\prime$.

\section{Physical Interpretation}
\label{sec:physics}

We have seen that the Faddeev integral~\eqref{define_Phi}  is closely
connected with the hyperbolic geometry.
Especially
we have shown that
classical limit of (imaginary part of) matrix elements  coincides with
the hyperbolic
volume of the ideal tetrahedron, and that the saddle point equation
coincides with the hyperbolic gluing condition.
As the 3-dimensional hyperbolic geometry is nothing but the Euclidean
AdS space,
we recall here the so-called AdS/CFT correspondence;
gravity theory on a 3-dimensional anti-de Sitter space is equivalent to
the  conformal field theory on the boundary.
This was initiated from the observation~\cite{BrowHenn86a}
that the asymptotic symmetry group of AdS$_3$ is generated by left and
right Virasoro algebra.
In this correspondence the CFT lives on the cylindrical boundary of
AdS$_3$, and in our viewpoint 
it seems that this cylindrical boundary can be
identified with a knot $K$.

To clarify   our observation  further,
it is preferable  to recall
properties of the
Faddeev integral~\cite{LFadd95a,FaddKashVolk00a}.
Collecting  residues in  integral in eq.~\eqref{define_Phi},
we get
(we assume $\Im \gamma >0$)
\begin{equation}
  \label{define_phi_exp}
  \Phi_\gamma(\varphi)
  =
  \frac{E_q(\mathrm{e}^\varphi)}{
    E_Q(\mathrm{e}^{\varphi \pi / \gamma})
    } ,
\end{equation}
where
\begin{align*}
  q
  & = \mathrm{e}^{\mathrm{i} \gamma} ,
  &
  Q
  &=\mathrm{e}^{-\mathrm{i} \pi^2/\gamma} ,
\end{align*}
and $E_q(w)$ is the $q$-exponential function defined by
\begin{align}
  \label{q_exponential}
  E_q(w)
  & =
  \prod_{n=0}^\infty (1 + q^{2 n + 1} \, w) .
\end{align}
With this function
the five-term relation~\eqref{pentagon_Phi} is replaced
by~\cite{FaddKash94}
\begin{gather}
  \label{pentagon_E_q}
  E_q(\Hat{b}) \, E_q(\Hat{a})
  =
  E_q(\Hat{a}) \, E_q(q^{-1} \, \Hat{a} \, \Hat{b}) \,
  E_q(\Hat{b}) ,
\end{gather}
with the Weyl commuting operators
\begin{gather*}
  \Hat{a} \, \Hat{b} = q^2 \, \Hat{b} \, \Hat{a}  .
\end{gather*}
With a definition of a  parameter $q$,
and setting
$\Hat{a}=\exp \Hat{q}$ and $\Hat{b} = \exp \Hat{p}$,
we can see from eq.~\eqref{canonical} that we  can naively replace
the Faddeev integral $\Phi_\gamma(w)$ in eq.~\eqref{represent_S} with
$E_q(\mathrm{e}^w)$
as was  shown in eq.~\eqref{define_phi_exp}.
In fact asymptotic behavior of the $q$-exponential is same with that of
$\Phi_\gamma(\varphi)$.
We can check using the Euler--Maclaurin formula
that
the classical limit $\gamma\to 0$ of  $E_q(\mathrm{e}^w)$  indeed
reduces to the Euler dilogarithm function as in
eq.~\eqref{Phi_and_Euler}
as follows;
\begin{align*}
  \log E_q(w)
  & =
  \sum_{n=0}^\infty \log ( 1 + \mathrm{e}^{\mathrm{i} \gamma (2n+1)}
  \, w)
  \\
  & \sim
  \frac{1}{2 \,  \mathrm{i}\, \gamma}
  \int_{-w}^0 \frac{\log(1-s)}{s} \mathrm{d} s
  =
  \frac{1}{2 \, \mathrm{i} \, \gamma} \,
  \Li(-w).
\end{align*}
Note that remaining part, $E_Q(\mathrm{e}^{\Hat{p} \pi/\gamma})$,
also satisfies the five-term relation~\eqref{pentagon_E_q}
replacing $q$ by $Q$, and
that
the vertex operator
$\mathrm{e}^{\Hat{p}}$
commute with a dual vertex operator
$\mathrm{e}^{\Hat{p}\pi/\gamma}$.
This type of  modular invariance,
$\gamma \leftrightarrow \pi^2/\gamma$,
can also be seen in CFT.
In this sense the integral~\eqref{define_Phi} is a dualization of the
$q$-exponential function.



\begin{figure}[htbp]
  \begin{center}
    \begin{psfrags}
      \psfrag{s}{$\sigma^2$}
      \psfrag{t}{$\sigma^1$}
      \psfig{file=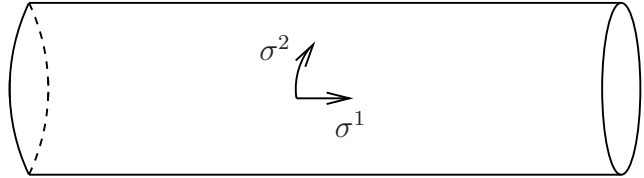,scale=0.4}
    \end{psfrags}
    \caption{Cylinder in a limit $\gamma\to 0$.}
    \label{fig:cylinder}
  \end{center}
\end{figure}

As a result,
matrix element,
$\langle p_1, p_2 \ | \ S^{-1} \ | \
p_1^\prime , p_2^\prime \rangle$,
defined
in eq.~\eqref{represent_S}
can be rewritten in terms of a function  defined by
\begin{gather}
  \label{define_boson}
%
  Z_{\text{B}}(\mu,N)
  =
  \mathrm{e}^{\frac{1}{2 \mathrm{i}  \gamma} (C- \mu N)}
  \,
  \frac{1}{
    E_q(-q^{-1} \mathrm{e}^\mu)
    } ,
\end{gather}
where
we set $C=\frac{\pi^2 + \gamma^2}{6} + \frac{\gamma \pi}{2}$.
The function $Z_{\text{B}}(\mu,N)$ is familiar in CFT, and
coincides with
the partition function of the free
boson up to constant.
In this view,
$\mu$ and $N$ correspond to the chemical potential and the
scaled average number of angular momentum respectively.
The constant term $C$   cancels if we introduce  
super partner
whose partition function can be written as
\begin{gather}
  \label{define_fermi}
%
  Z_{\text{F}}(\mu, N)
  =
  \mathrm{e}^{\frac{1}{2 \mathrm{i} \gamma} ( -C + \mu N)}
  \,
  E_q(-q \, \mathrm{e}^\mu) .
\end{gather}
This constitutes to matrix element,
$  \langle p_1 , p_2 \ | \ S \ | \
p_1^\prime , p_2^\prime \rangle$
in eq.~\eqref{represent_S}.
It should be noted that
the above  functions $Z_{\text{B},\text{F}}(\mu,N)$
can be identified with the partition functions of bosonic (fermionic)
string from the cylinder,
and that
the classical limit $\gamma\to 0$
corresponds to
a  long  strip  limit
(Fig.~\ref{fig:cylinder}).

It is now clear that,
by replacing $\Phi_\gamma(\varphi)$ in  eq.~\eqref{represent_S}
with the $q$-exponential function~\eqref{q_exponential},
the   bosonic  string
corresponds to the oriented tetrahedron
(say, left in  Fig.~\ref{fig:tetrahedron})
while
its super-partner
to  mirror image
(say, right in Fig.~\ref{fig:tetrahedron}).
It seems that the  quantum five-term relation~\eqref{pentagon_Phi}
indicates  a possibility to construct
the quantum invariant~\eqref{partition_function}
by using
\emph{noncommutative}
thermodynamic  variables $\mu$ and $N$ as in eq.~\eqref{canonical}.
In this formalism,
string interactions should be regarded as
constraints among $\mu$
and $N$,
and in the classical limit they  describe
the hyperbolic consistency condition.

To conclude the partition
functions~\eqref{define_boson}--~\eqref{define_fermi} can be
seen to be
fluctuating around
the oriented ideal tetrahedron in the Euclidean AdS$_3$.
This fact seems to indicate  aspects of the AdS/CFT correspondence.


\section{Concluding Remarks}

We have studied the hyperbolic structure of
the partition function which is defined by use of
the quantum
dilogarithm function (Faddeev integral).
We have shown  that
the classical limit of the Faddeev integral
describes the
ideal
tetrahedron in the hyperbolic space $\mathbb{H}^3$ which 
is the
space-like surface in 4-dimensional Minkowski space.
Though we have originally defined the partition function in terms of
the Faddeev integral, we can replace it with the $q$-exponential
function as we have discussed in Section~\ref{sec:physics}.
As the $q$-exponential function
denotes a partition function of free bosonic and fermionic strings
from cylinder,
we have proposed 
that the hyperbolic structure $\mathbb{H}^3$
can be 
naturally embedded into (classical limit of)
the CFT, and that it indicates
a new aspect of the  AdS/CFT
correspondence.
As a result,
the modulus of the oriented ideal tetrahedron
which is assigned to partition functions of
free boson and fermion,
is given  explicitly in terms of the
chemical potential and the average number of angular momentum, and
the string interaction should be regarded as a quantization of the
hyperbolic consistency conditions in gluing ideal tetrahedra.
We note that we have assigned the $q$-exponential function to
hyperbolic ideal tetrahedron to construct the quantum
invariant~\eqref{partition_function}, while 
an idea of the Regge calculus~\cite{PonzaTRegg68a} is 
that the three dimensional gravity is a sum over $6j$ symbol.

Promising is a fact
that the classical limit~\eqref{volume_conjecture}  of the partition
function $\tau(M)$  is suggested to give
the hyperbolic volume and the Chern--Simons
invariant~\eqref{represent_VCS} of manifold $M$.
As it was shown~\cite{EdWitt88d}
that the Einstein--Hilbert action with negative cosmological constant
$\Lambda=-1/\ell^2$,
\begin{equation*}
  S = \frac{1}{16 \, \pi \, G}
  \int \mathrm{d}^3 x
  \sqrt{-g} \,
  \Bigl(R + \frac{2}{\ell^2}\Bigr)
\end{equation*}
reduces to 
2 sets  of the  Chern--Simons actions
by taking a linear combination of
the spin connection and the vierbein as the gauge field,
the partition function $\tau(M)$ 
seems to a candidate for quantum gravity.

The AdS/CFT correspondence  enables us to derive the
Bekenstein--Hawking entropy of black hole
microscopically~\cite{StroVafa96a}.
Therein it was shown that
the entropy coincides with 
the asymptotic growth of the number of states in CFT with central
charge $c$~\cite{JLCard86b}.
We see that the computation of the central charge $c$ is
essentially same with our computation of the classical limit of the
partition function~\eqref{volume_conjecture},
once
we
have identified   the integral $\Phi_\gamma(\varphi)$ with
the  partition function of free boson and
fermion written by
the $q$-exponential function.
As we have seen based on the ``volume conjecture'' that
the asymptotic behavior determines  the hyperbolic geometrical
structure, 
it would help us to understand black hole.


\appendix
\section{Dilogarithm Function}

\begin{itemize}
\item 
  the Euler dilogarithm,
  \begin{equation}
    \label{define_Euler}
    \Li(z)
    =\sum_{n=1}^\infty \frac{z^n}{n^2}
    = - \int_0^z \mathrm{d} s \
    \frac{\log(1-s)}{s} 
  \end{equation}

\item the Rogers dilogarithm,
  \begin{equation}
    \label{define_Rogers}
    L(z)
    = \Li(z) + \frac{1}{2} \log(z) \log(1-z) 
  \end{equation}
\item the Bloch--Wigner function,
  \begin{equation}
    \label{define_Bloch_Wigner}
    D(z)
    = \Im \, \Li(z) +
    \arg(1-z) \cdot \log|z| 
  \end{equation}
\end{itemize}

\newpage


\end{document}